\documentstyle[twocolumn,prc,aps,epsf]{revtex}

\begin{document}

\title{NEW MECHANISM FOR THE PRODUCTION OF THE EXTREMELY FAST LIGHT PARTICLES
 IN HEAVY-ION COLLISIONS\\ IN THE FERMI ENERGY DOMAIN\thanks{Proceeding of VII
 International School-Seminar on Heavy-Ion Physics, May 27 -- June 1, 2002, Dubna, Russia.}}
\author {A.S. Denikin\thanks{E-mail: denikin@jinr.ru}, V.I. Zagrebaev}
\address{141980, Joint Institute for Nuclear Research, Dubna, Moscow region, Russia}
\date{August 21, 2002}
\maketitle

\begin{abstract}
Employing a four-body classical model, various mechanisms responsible for the production
of fast light particles in heavy ion collisions at low and intermediate energies have
been studied. It has been shown that at energies lower than 50 A MeV, light particles of
velocities of more than two times higher than the projectile velocities are produced due
to the acceleration of the target light-particles by the mean field of the incident
nucleus. It has also been shown that precision experimental reaction research in normal
and inverse kinematics is likely to provide vital information about which mechanism is
dominant in the production of fast light particles.
\end{abstract}

\section{Introduction}

The production of pre-equilibrium light particles (n, p, t, $\alpha $) in nucleus-nucleus
collisions depends on the way the nuclear system evolves at the reaction initial stage.
For heavy ion collisions, the light particle cross section is known to be a noticeable
fraction of the total reaction cross section even at low energies of the order of 10 A
MeV, i.e. light particle production is a distinctive feature of all nuclear reactions
involving heavy ions. This means that studying the production mechanisms for those
particles is likely to provide direct information on both the reaction initial stage
dynamics and the potential and dissipative forces of nucleus-nucleus interaction.

\begin{figure}[t] \epsfxsize=7.5 cm
\centerline{\epsfbox{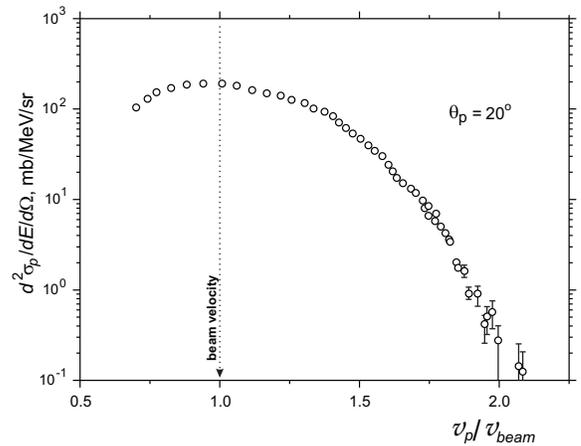}} \vskip 0.1 cm

\caption{ Measured differential cross section of protons emitted at the angle
$\theta_{lab} = 20^\circ$ in the reaction $^{16}O + ^{197}Au \to p + X$ at the energy 20
A MeV plotted versus the proton velocity in units of the beam velocity. The experimental
data are taken from [1].} \label{Fig1}
\end{figure}

Vast bodies of data \cite{Awes,Sackett,Lanzano,Penion,Alba,Sapienza} demonstrate that in
heavy ion reactions at energies per nucleon of the order of the Fermi energy, light
particles are produced in a wide angular range, their velocities being two and more times
higher than the velocities of the beam particles. Fig. \ref{Fig1} shows the measured
differential proton cross section at $\theta _{lab}  = 20^ \circ $ in the case of the
${}^{16}O$ (20 A MeV) + ${}^{197}Au$ collision \cite{Awes}. What is the mechanism of the
production of these extremely fast light particles? The ultimate answer to that question
has not been provided yet. Circumstantial evidence suggests that they are of a
pre-equilibrium nature. This makes them more difficult to study by a direct experiment
since there is currently no way of studying processes taking place in time intervals of
the order of $10^{ - 20} $ s.

Attempts to interpret the experimental picture have resulted in the creation of a number
of theoretical models and approaches. Among these in particular are the models of moving
sources \cite{Awes2}, Fermi-jet \cite{Bondorf,Mohring,Randrup}, dissipative break-up and
massive transfer \cite{Zagr} etc. A comprehensive survey of experimental and theoretical
works relating to this problem is given in \cite{Zagr2}. Applying these approaches, the
authors have succeeded in qualitatively describing the energy and angular dependencies of
the spectra of emitted light particles as well as revealing some production mechanisms
for them. Among the latter in particular is the mechanism of "splashing" nucleons out of
a retarding projectile nucleus, the high velocities of the light particles being due to
the addition of the velocity of Fermi motion within the projectile nucleus and the
velocity of its centre of mass. The importance of taking this mechanism into account was
shown in refs. \cite{Mohring,Zagr}, when describing the spectra of the $\alpha$-particles
produced in projectile break-up and massive transfer at collision energies of the order
of several tens of MeV per nucleon.

Nucleon-nucleon collisions are also likely to result in a fast light particle being
emitted due to scattering on nucleons of high velocities. The Fermian shape of nucleon
momentum distributions explains the existence of such nucleons in a heated nuclear
system. The first mechanism, capable of qualitatively describing the experimental results
in a transparent way, fails to provide satisfactory quantitative agreement. The
nucleon-nucleon collision model is adequate at high energies ($<$ 100 A MeV), but in the
region of energies $E_0 <$ 50 A MeV the nucleon-nucleon interaction cross section for
heavy ion collisions decreases due to Pauli's principle. Therefore this mechanism
influences the formation of the hard part of the spectrum of light particles to a smaller
degree. Recent advances in instrumentation have made possible the precise measurement of
the angular and energy spectra of light particles \cite{Penion,Alba,Sapienza}, which is
likely to shed light upon yet unresolved problems.

\section{Mechanisms of Light-Particles Formation}
In ref. \cite{Denikin}, we proposed a four-body classical model of nucleus-nucleus
collisions, which permits establishing the role of mean nuclear fields and
nucleon-nucleon collisions in the production of light-particles. This model considers the
projectile and target nuclei to be bound two-body systems that are composed of a heavy
core and a light fragment moving in its field. The interactions of the fragments, which
follow classical trajectories, define six pair potentials. The interaction between the
light fragments and the cores was described by the Woods-Saxon potential with
compilation-based parameters \cite{Perey}. The interaction between the heavy cores was
described by the proximity potential or the Woods-Saxon potential, the parameters of the
latter being chosen so as to reproduce the height and position of the Coulomb barrier.
The coupling with the reaction channels, which were not taken into account explicitly,
was introduced by means of dissipative forces acting between the heavy cores of the
target and projectile. The choice of friction coefficients and form-factors for the
dissipative forces was based on ref. \cite{Gross}. To provide a more correct estimate of
the channel differential cross sections, the description of the relative motion of the
projectile (target) light particle and the target (projectile) core has an absorption
probability for them has been introduced as $P_{ij}^{abs}  = 1 - \exp \left( { -{{s_{ij}}
\mathord{\left/{\vphantom {{s_{ij} } {\lambda _{ij}}}} \right. \kern-\nulldelimiterspace}
{\lambda _{ij} }}} \right)$, where $s_{ij}$ is the distance traveled by particle i in the
nucleus j, $\lambda _{ij} $ – the corresponding free path length, which is, as well
known, related to the imaginary part of the optical potential, the parameters of which
were chosen according to ref. \cite{Perey}. Specifying the relative distance vector ${\bf
r}_{ij} $ for the projectile components, the binding energy $E_{ij}^{sep} $ , known by
experiment, as well as the orbital momentum $l_{ij} $ for their relative motion,
completely defines the projectile inner spatial configuration. The components of the
vector ${\bf r}_{ij} $ were chosen randomly and equiprobably in the classically allowed
region. To have the relative motion momentum ${\bf p}_{ij} $ uniquely determined, one of
the components of the vector ${\bf l}_{ij} $ must also be specified (by a random choice)
in addition to the $r_{ij} $ and $E_{ij}^{sep} $ values. Repeating the same operations
for the target nucleus and specifying the relative motion of the centers of mass of the
nuclei depending on the reaction allow to determine the initial conditions that are
required for solving the set of classical equations of motion.

\begin{figure}[t]
\epsfxsize=7.5 cm \centerline{\epsfbox{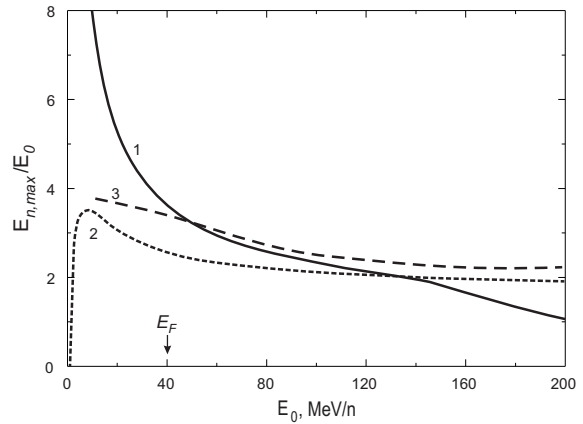}} \vskip 0.1 cm \caption{The ratio of
maximum energy $E_{n,\max }$ of pre-equilibrium neutron to the beam energy $E_0 $
calculated for the reaction ${}^{20}Ne$ + ${}^{165}Ho \to $ n + X as a function of the
beam energy. The solid line corresponds to the maximum energy of neutrons originating
from target, the short-dashed line – to the maximum energy of neutrons emitted from the
projectile, and the dashed line – to the maximum energy of neutrons emitted after
nucleon-nucleon collisions. The arrow represents the beam energy per nucleon
approximately equal to the Fermi energy.} \label{Fig2}
\end{figure}

The model outlined above has 15 reaction channels with a different combination of
particles in the final fragments. Eight of these channels contribute to the total cross
section for pre-equilibrium light particles, which can be divided into two groups:
particles emitted from the projectile and from the target. Taking into account all the
pair potentials allows us to study the whole range of processes that result in
light-particle emission as well as to establish the role played in them by one or another
type of interaction. We have investigated heavy ion collisions at energies of several
tens of MeV per nucleon and thoroughly studied the production mechanisms for fast light
particles. It is remarkable and unexpected that the hardest part of the energy spectrum
of emitted light particles corresponds to particles emitted from the target nucleus
rather than from the projectile nucleus, as has been assumed up to now. There is a simple
enough explanation for this phenomenon, which appears to be unusual at first sight. Let
us assume for simplicity that the heavy cores have masses much larger than those of the
light particles; the light particle moves in the nucleus at a velocity equal to the Fermi
velocity $v_F $ ; the nuclei move relative to one another at the velocity $v_0  \approx
v_F $ ; the trajectories of the heavy fragments are taken to be undistorted; the binding
energy of the light particle in the nucleus is assumed to be much smaller than the
collision energy $E_{ij}^{sep}  \ll E_0  = {{mv_0^2 } \mathord{\left/{\vphantom {{mv_0^2
} 2}} \right. \kern-\nulldelimiterspace} 2}$, here m is the mass of the light particle.
Then, as has been shown in [13], for example, the neutrons emitted from the projectile
("splashing out") will have a maximum laboratory system energy expressed as $$E_{n,\max }
\approx \frac{m}{2}\left( {v_0^2  + 2v_0 v_F } \right) = 3\frac{{mv_0^2 }}{2} = 3E_0.
\eqno{(1)}$$

In the framework of this model under the same assumptions, the maximum energy of the
neutrons emitted both from the projectile and from the target in elastic nucleon-nucleon
collisions is limited by the value $$ E_{n,\max }  \approx \frac{m}{2}\left( {v_0  + v_F
} \right)^2 = 4\frac{{mv_0^2 }}{2} = 4E_0. \eqno{(2)}$$ However, such a mechanism of
emitting a fast nucleon suggests that the recoil nucleon is to impart the whole of its
kinetic energy to the particle knocked out and, consequently, pass into an occupied lower
energy state, which is forbidden by Pauli's principle. Therefore, this mechanism does not
manifest itself in full measure at low and intermediate collision energies and comes into
play as the collision energy increases, when states more and more lower in energy become
unoccupied in the process of the system getting excited.

The production mechanism for the fast light particles emitted form the target is more
complicated. To begin with, let us consider the elastic scattering of a target light
particle on a moving infinitely heavy repulsing wall. If before colliding, the light
particle and the heavy wall have been moving collinearly in opposite directions with
parallel velocities $v_F $ and $v_0 $ respectively, then the centre-of-mass velocity of
the light particle has been $v_{cm} = - (v_0  + v_F )$ . After elastically colliding, the
light particle will have the velocity $v'_{cm}  =  - v_{cm} = (v_0  + v_F )$ , the lab
velocity being $v' = v'_{cm}  + v_0  = 2v_0  + v_F $ . Thus the maximum velocity of a
neutron emitted from the target will be $$ E_{n,\max }  \approx 4\frac{m}{2}\left( {v_0^2
+ v_0 v_F } \right) = 8\frac{{mv_0^2 }}{2} = 8E_0, \eqno{(3)}$$ i.e. the fastest neutrons
will have a velocity of more than 2.5 times higher than that of the beam particles. Now,
suffice it to say that an elastic collision with a repulsing wall is kinematically
equivalent to elastic scattering at the angle $\theta _{cm}  =  - 180^ \circ$ (i.e.
orbiting) in the attractive field of an incident nucleus. However, orbiting is possible
at quite low relative motion energies. At higher energies, a light particle can be
deflected only due to the field of the projectile at a certain centre-of-mass angle
$\theta _R $ called an angle of rainbow scattering. Therefore the maximum energy of a
pre-equilibrium light particle emitted will be a function of collision energy,
light-particle-incident-nucleus interaction potential, particle binding energy and
dissipative forces.

More accurate calculations of the maximum energy of the pre-equilibrium neutrons produced
in the reaction ${}^{20}Ne$ + ${}^{165}Ho \to $ n + X are presented in Fig. \ref{Fig2}.
Given here is the $E_{n,\max } /E_0 $ ratio as a function of the beam energy $E_0 $ for
the above mechanisms: the solid curve is for the energy of the neutrons emitted from the
target; the dashed curve for the energy of the neutrons emitted from the projectile; the
dash-dotted curve for the energy of neutrons produced in nucleon-nucleon collisions,
which was calculated with regard to Pauli's principle. So, it is seen that at collision
energies per nucleon smaller than $E_F $ (indicated with an arrow in Fig. \ref{Fig2}),
the mechanism involving acceleration of neutrons by the moving mean field of the
projectile is dominant in the formation of the high-energy portion of neutron spectra. In
the high-energy region, the key role is played by nucleon- nucleon collisions. It should
be also noted that the curves in Fig. \ref{Fig2} do not represent $E_{n,\max }$ in a
strict way since our calculations did not take into account the non-zero probability of
there existing nucleons of velocities much higher than $v_F $. Therefore the boundaries
indicated with the curves in Fig. \ref{Fig2} will shift to higher energies.

\begin{figure}[t]
\epsfxsize=7.5 cm \centerline{\epsfbox{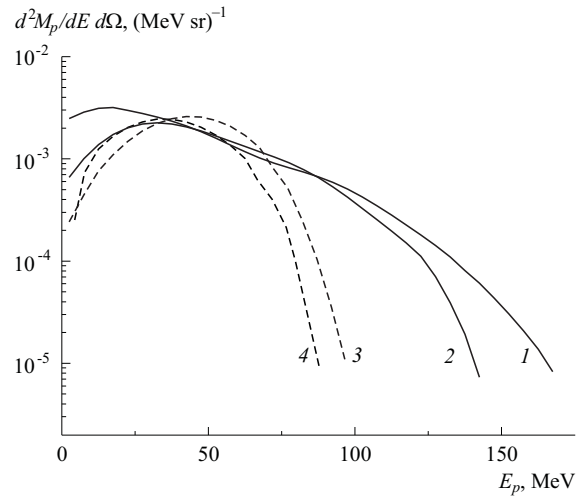}}\vskip 0.1 cm \caption{Calculated
differential multiplicity of pre-equilibrium protons emitted at the angle $\theta _{lab}
= 51^ \circ $ in the reactions ${}^{40}Ar$ + ${}^{51}V$ and ${}^{132}Xe$ + ${}^{51}V$ at
the incident energy 44 A MeV. Curves 1 and 2 correspond to the protons, originating from
the target ${}^{51}V$ bombarded with ${}^{132}Xe$ and ${}^{40}Ar$ projectiles,
respectively. Curves 3 and 4 correspond to the protons, emitted from the ${}^{132}Xe$ and
${}^{40}Ar$ projectiles, respectively.} \label{Fig3}
\end{figure}

How should the predictions, made by our model be tested, and how should the yields of
pre-equilibrium light particles emitted from a projectile and target be separated? There
is currently no direct solution. However the above analysis of various processes of
light-particle production in heavy ion collisions shows that the mechanism of emitting
fast nucleons from the target is only slightly sensitive to the projectile nucleus mass,
and, consequently, the maximum energy of light particles is not likely to change greatly
with replacing one projectile by another. On the other hand, the maximum energy of the
target nucleons accelerated by the attractive field of the incident nucleus is defined by
the corresponding rainbow scattering angle, which depends on the dimensions of the
deflecting mean field \cite{Knoll} $$\theta _R  \approx {\left(V_C  - 0.56 U_0
\sqrt{{R_U}\left/{a_U}\right.} \right)} \left/ {E_{cm}}\right., \eqno{(4)}$$ where $V_C $
is the height of the Coulomb barrier for the emitted light particle; $U_0 $, $R_U $, $a_U
$ – the depth, radius and diffusiveness of the light-particle-projectile interaction
potential, and $R_U  \approx 1.3A^{{1 \mathord{\left/ {\vphantom {1 3}} \right.
\kern-\nulldelimiterspace} 3}} $. It should be noted, the coefficient 0.56 in empirical
equation (4), obtained in ref. \cite{Knoll}, is not valid in case of nucleon scattering
on heavy nucleus. We found that coefficient 0.7 enables to obtain more reasonable value
of nucleon rainbow angle $\theta_R$.

Now it is evident that the energy distribution of emitted light particles must be more
elongated for a reaction with a heavy projectile than for a reaction with a light one.
This conclusion is supported by Fig. \ref{Fig3}, in which the calculated differential
pre-equilibrium proton cross sections are shown for the angle $\theta _{lab}  = 51^ \circ
$ for the ${}^{40}Ar$ + ${}^{51}V$ and ${}^{132}Xe$ + ${}^{51}V$ reactions (in normal and
inverse kinematics) at the energy $E_0$ = 44 A MeV. It is seen, first, that the key role
in the formation of the high-energy part of the spectrum is played by the protons emitted
from the target nuclei (curves 1 and 2 in Fig \ref{Fig3}). Second, the distribution of
the protons emitted from projectiles of different masses (curves 3 and 4 in Fig.
\ref{Fig3}) only changes its shape to a large degree at small energies, i.e. a more
massive ${}^{132}Xe$ projectile loses less energy in a collision with a ${}^{51}V$ target
than does ${}^{40}Ar$ . Therefore these two reactions have projectile proton
distributions with shifted maxima. Third, the ${}^{40}Ar$ induced reaction in normal
kinematics has a pre-equilibrium-proton spectrum, which is several tens of MeV shorter
than that for the reaction with the $^{132}Xe$ heavier projectile. So comparing precisely
measured data on the distributions of the light particles produced in heavy ion
collisions in reactions of normal and inverse kinematics will give us vital information
about which mechanism is dominant in emitting a fast light particle.

\end{document}